\documentclass[runningheads]{llncs}
\usepackage{graphicx}
\usepackage[colorinlistoftodos]{todonotes}

\begin{document}
\title{Usage-based Summaries of  Learning Videos\thanks{The research was partly supported by Science Foundation Ireland under Grant Number SFI/12/RC/2289\_P2, co-funded by the European Regional Development Fund.}}

 \author{Hyowon Lee\orcidID{0000-0003-4395-7702} \and Mingming Liu \orcidID{0000-0002-8988-2104} 
 \and \\Michael Scriney \orcidID{0000-0001-6813-2630} \and Alan F. Smeaton\orcidID{0000-0003-1028-8389}}

\authorrunning{H. Lee et al.}
%
\institute{Insight Centre for Data Analytics, Dublin City University,  Dublin 9, Ireland\\
\email{Alan.Smeaton@dcu.ie}
}
\maketitle              
\begin{abstract}
Much of the delivery of University education is now by synchronous or asynchronous video.  For students, one of the challenges is managing the sheer volume of such video material as  video presentations of taught material are difficult to abbreviate and summarise because they do not have highlights which stand out. Apart from video bookmarks there are no tools available to determine which parts of video content should be replayed at revision time or just before examinations.   We have developed and deployed a digital library for managing video learning material which has many dozens of hours of short-form video content from a range of taught courses for hundreds of students at undergraduate level. Through a web browser we allow students to access and play these videos and we log their anonymised playback usage.  From these logs we score to each segment of each video based on the amount of playback it receives from across all students, whether the segment has been re-wound and re-played in the same student session, whether the on-screen window is the window in focus on the student's desktop/laptop, and speed of playback.  We also incorporate negative scoring if a video segment is skipped or fast-forward, and overarching all this we include a decay function based on recency of playback, so the most recent days of playback contribute more to the video segment scores.  For each video in the library we present a usage-based graph which allows students to see which parts of each video attract the most playback from their peers, which helps them select material at revision time.  Usage of the system is fully anonymised and GDPR-compliant. 

\end{abstract}

\section{Pedagogical Background}

University classes virtually conducted on Zoom or other online platforms as the result of COVID-19 have had an  immediate impact on how  students study and review what was delivered during the semester in preparation for semester-ending exams. For classes delivered synchronously (e.g. an online, live lecture over Zoom), the recording of those sessions is useful and thus typically is made available on a learning management systems (LMS) for  students to re-watch later on; for classes that use asynchronous video materials (e.g. a series of short video screencasts in which a lecturer may explain concepts), typically many short or long recorded videos become available for students to watch in their own time. Many university courses have been employing a mixture of these synchronous and asynchronous methods to compensate for the lack of the benefits of face-to-face class setting. Students' views on using such educational videos as the main source of learning has recently been studied \cite{Krieter2020} showing a mixture of benefits and fears.

Partly due to the inherent temporal nature of video medium that requires a viewer to ``playback'' in order to understand the contents, and partly due to the unedited and linear nature of lecture videos (unlike production videos featuring content-induced structure such as camera shots and scenes, chapters, and transitions), one consequence in the consumption of the video materials generated in this context is an increased burden to the students who face a large amount of unstructured lecture videos and screencasts at the time of reviewing: simply re-watching all video materials is not feasible, and yet there is no way to know what parts of the videos they should focus on. There are recent studies on tracking students' eye-gaze while watching a video lecture \cite{Srivastava2021} or logging their level of attention during online classes \cite{Lee2021}, the knowledge of which could be used in suggesting the parts of videos that  students should/could focus on in reviewing. This requires the capturing of the data at the time of watching/lecturing, an overhead on the front-end such as camera or other installed software to capture the eye-tracking/attention data.

The system described in this demonstration paper is a web-based video library of recorded video materials (both synchronous and asynchronous) as a result of running a remote online course at our university. By recording and analysing the detailed playback usage of each video including fast-forward, jumping forward while playing and re-playing over time by the student cohort, the system visualises and highlights which portions within each video have been found to be most important or most used by other  students, thus offering clues during the re-watching/ revision process without requiring any manual intervention (e.g. lecturer indicating the parts of videos to watch).

\section{Description of the Prototype}

Our system generates a usage visualisation purely based on the playback-related interaction logs incurred by anonymised students, after the videos are made available for them for viewing on the course. The visualisation is a time-based graph aligned with the playback timeline of video content, where the height along the timeline indicates usage scores calculated using the strategy below. 
The playback-related interactions captured and factored in by the system include playing/pausing, seeking/skip forward (i.e. jumping from one point to another within a video),  video playback window moving in/out of focus, and playback speed/rate.

Each video is divided into 1-second windows and each window starts with an initial score of 0 which is incremented every time any student plays or skips it. Every time any part of the video is played, that part will gain a score, thus over time as the usage increases the score in each second-by-second window on the timeline will increase. An important consideration is  how much score gain each 1-second-window should receive from each of the playback-related interactions above, in order to result in a meaningful and useful indication when accumulated over time, in guiding students in their selective watching.

Our strategy and the rationale for calculating the score gain for each 1-second-window is summarised below:

\begin{itemize}
\item \emph{Playback as part of ``run through''}: As the most basic scoring strategy, the window gets \textbf{+1} when that portion is played. However, if the video playback window was not the window of focus on the student's screen when the student was playing it, then this increment is \textbf{+0.25} only (the student may be reading email or something else while half-listening to the video).

\item \emph{Replay as part of rewind within the same session} will gain \textbf{+2} score for each 1-second-window so it gets a cumulative +3: +1 from the initial playback and +2 from the replay). This assumes a replay is done with the playback window as the window in focus, as it would not be possible to select and replay if the window was not the focus.

\item \emph{Playback at $2\times$ (double speed)} will gain \textbf{+0.6} and if the window is not the window of focus then it will be \textbf{+0.2}  only because when the student is attending another window, the double-speed playback is too fast to properly comprehend.

\item \emph{Playback at $1.5\times$} will gain \textbf{+1.5}, and \textbf{+0.5} if the playback window is not the window of focus. In fact, a moderate acceleration of the video playback may potentially lead to increased students' learning performance. For instance, a recent study \cite{10.1145/3375462.3375466} shows that playing  educational videos at $1.25\times$ resulted in better outcomes (e.g. getting higher grades) than normal speed, while also testing for other speeds ($0.75\times$ and $1.75\times$).

\item \emph{Skip forward}: If a student skips forward from the current position at minute S0 by a segment of video then windows within  1-, 2-, and 3-minute  segments directly following the  segment S0 will get score adjustments as follows: S60: \textbf{-0.3}, S120: \textbf{-0.2}, and S180: \textbf{-0.1}. The rationale for this deduction is that the student must have had an idea what was coming up next, after S0, but anticipated it as being not interesting or useful thus less likely for other students to find it also  interesting or useful. 

\item \emph{Adjusting the score over time}: Overarching  this  scoring strategy is a \textbf{decay function} based on recency of playback, with the most recent days of playback interaction being more meaningful or useful than prior to that. The scores calculated by the above strategy are re-calculated from the interaction log each midnight. In this way the score increments (both + and -) as above are called our BaseIncrements = (+1.0, +0.25, +2, +0.6, +0.2, +1.5, +0.5, -0.3, -0.2, -0.1) and the system makes those the actual increments on day 0. Then on day 1 it makes those increments each multiplied by 1.1  before adding, on day 2 the base increments  multiplied by 1.2, and so on.  The effect is that on day 10 we have a score for each 1-second window which has a 10-day linear decay function so that something played on day 9 has twice the value of something played on day 0 and by day 20 we have a score which has a 20-day linear decay with the ``half-life'' being 10 days.  This  continues  indefinitely.
\end{itemize}

\begin{figure}[ht]
    \centering
    \includegraphics[width=0.85\textwidth]{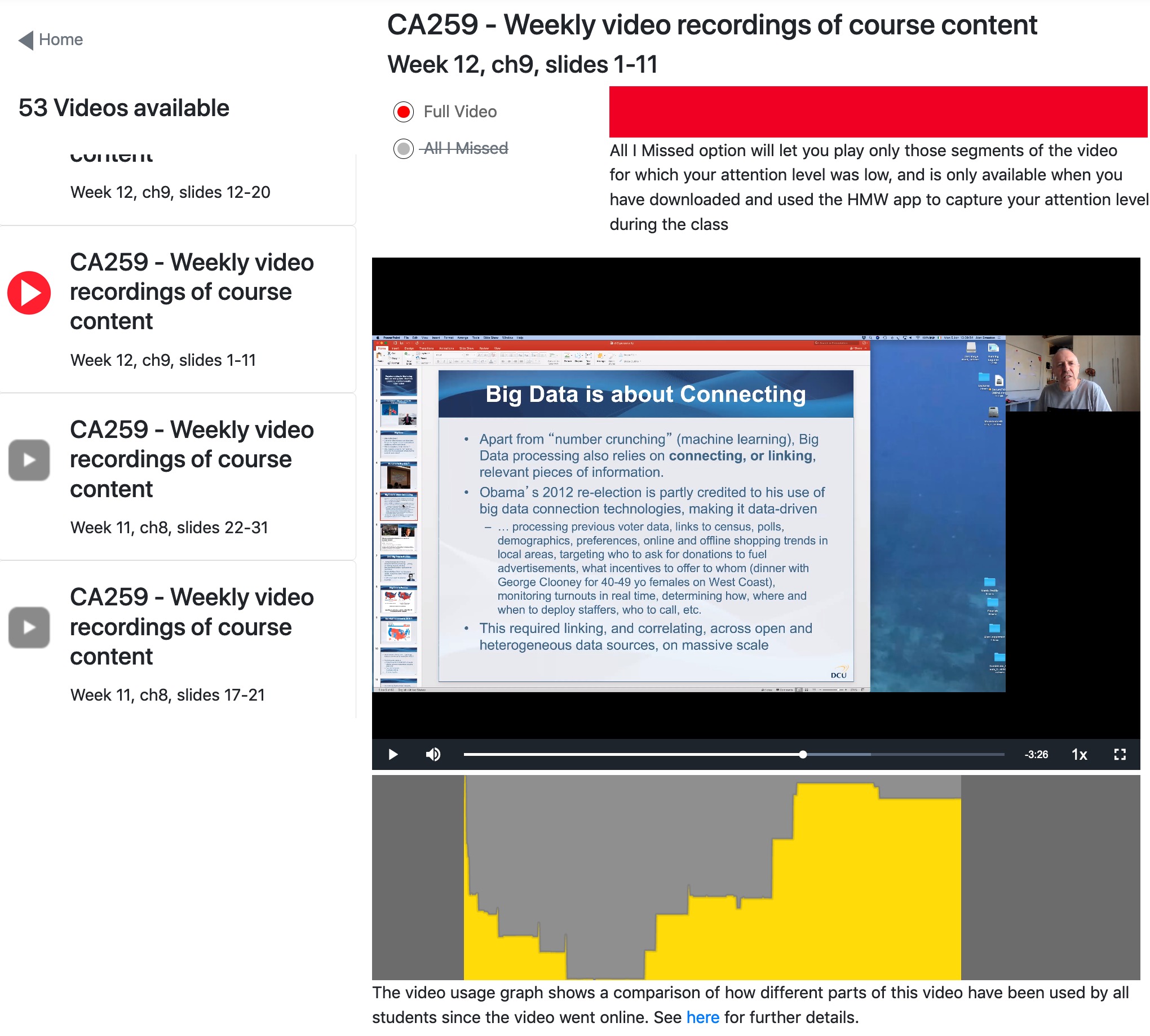}
    \caption{Screenshot of system interface showing the user has 3:26 left to play at 1x speed of what appears as a 10-minute video from week 12, chapter 9, slides 1 to 11 of course CA259. The yellow graph under the video playback window indicates the section the user is about to play has had highest usage rating based on previous video playback from the class whereas the part of the video at about the 1/3 mark has near zero usage.}
    \label{fig:screenshot}
    \label{fig:my_label}
\end{figure}

\noindent
The scores for the 1-second windows are normalised within each video usage graph at display time making the visualisation less susceptible over time to any maligned attempt at artificially inflating scores by jumping to or repeatedly playing an obscure segment of video.  Figure~\ref{fig:screenshot} shows a screenshot of the system.

\section{Use Case and Results Achieved}

The system has been deployed in our university and we use as an example,  an undergraduate-level course during the Spring semester 2021, with a class size of 131 students. This has 11 synchronous class recordings of about 1 hour and 20 minutes each and 53 asynchronous short-form video screencasts of about 10 minutes each, amounting to approximately 23 and a half hours of playable video content. Since the semester started, students have been using the system, actively playing and re-playing the video contents as part of their studies thus feeding into the playback usage analysis which, in turn, helps guide the portions of each video for them to watch. 

At the time of submission, the system is being used extensively by students with 2,900 hours of video streamed across 434 distinct sessions. On average per-session each student watched 1.5 videos and spent 6.7 minutes (404.8 seconds) viewing materials. As a percentage of a video viewed per-session each student on average viewed 35\% with a standard deviation of 0.4. However as the semester is incomplete at the time of submission, and we expect a large increase in usage as we get closer to end-of-semester examinations, it is too early to draw any further conclusions from this usage.

\section{Future Work}

Since this is the first time the system has operated for the full duration of a semester, we are learning  from the usage data and how to leverage it. We plan to fine-tune the scoring strategy based on the gained insights as well as diversifying to a greater range of interaction logs (e.g. playback volume). The timeline visualisation will be further refined to more effectively convey the usage data. More formal usability testing with students will also take place. Further validation on the effectiveness of our system by using semantic feedback from both students and lecturers on video content will also be part of our future work.



%
%
%
\bibliographystyle{splncs04}
\bibliography{HMWectel}
%
%
%
%
%
\end{document}